\documentclass[11pt,a4paper]{emulateapj}

\usepackage{epsfig}
\usepackage{amsmath}
\usepackage{natbib}

\begin{document}

\title{The R\O mer Delay and Mass Ratio of the sdB+dM Binary 2M 1938+4603 \\ from \textit{Kepler} Eclipse Timings}

\author{Brad N. Barlow, Richard A. Wade, and Sandra E. Liss}
\affil{Department of Astronomy and Astrophysics, 525 Davey Lab,
  The Pennsylvania State University, University Park, PA 16802, USA}

\submitted{Received 2011 December 20; accepted 2012 April 12; published 2012 June 19}

\begin{abstract}
  The eclipsing binary system 2M 1938+4603 consists of a pulsating hot subdwarf B
  star and a cool M dwarf companion in an effectively circular
  three--hour orbit.  The light curve shows both primary and secondary
  eclipses, along with a strong reflection effect from the cool
  companion.  Here we present constraints on the component masses and
  eccentricity derived from the R\o mer delay of the secondary eclipse.  Using six months of
  publicly-available \textit{Kepler} photometry obtained in Short
  Cadence mode, we fit model profiles to the primary and secondary
  eclipses to measure their centroid values.  We
  find that the secondary eclipse arrives on average 2.06 $\pm$ 0.12 s
  after the midpoint between primary eclipses.  Under the assumption 
  of a circular orbit, we calculate from this time delay a mass ratio of q = 0.2691 $\pm$ 0.0018 and
  individual masses of M$_{\rm sd}$ = 0.372 $\pm$ 0.024 M$_{\sun}$ and M$_{\rm c}$
  = 0.1002 $\pm$ 0.0065 M$_{\sun}$ for the sdB and M dwarf, respectively.
  These results differ slightly from those of a
  previously-published light curve modeling solution; this difference, however, may
  be reconciled with a very small eccentricity, $e \cos \omega \approx$ 0.00004.  We also report an orbital period decrease
   of $\dot{P}$ = (-1.23 $\pm$ 0.07) $\times$ 10$^{-10}$.
  

\end{abstract}


\section{Introduction}

Around ten eclipsing hot subdwarf B (sdB) star binaries are known
presently (e.g., \citealt{for10}); they have orbital periods from 2-4
hours, and the companions are typically low-mass M dwarfs. Although
these subdwarfs outshine their companions at all visible (and most
infrared) wavelengths, the high albedo of the M dwarf and its large
subtended solid angle as seen from the sdB star lead to a strong
reflection effect, and, consequently, these systems exhibit both primary
eclipses (when the M dwarf transits the sdB) and secondary eclipses
(when the sdB blocks reflected light from the M dwarf).  Binary
population synthesis (BPS) models (e.g., \citealt{han02,han03,cla12}) show
they are likely a product of a past common-envelope (CE) stage during which
the sdB progenitor filled its Roche lobe while on the red giant branch
and lost most of its outer H envelope.  Assessing the distribution of
their orbital parameters and masses can test binary evolution scenarios
and help constrain the parameterizations in BPS codes.  Masses have been
derived from light-curve modeling solutions for the majority of known
detached sdB+dM systems (e.g., \citealt{for10}), but the accuracy of
these results has not been tested thoroughly using independent
techniques.

Precise measurements of the eclipse timings in sdB+dM binaries provide
an opportunity to measure the component masses in a relatively
model-independent way.  An observer watching one of these systems
face-on from a great distance would see a syzygy of the sdB, its
companion, and Earth every half orbital period, if the orbit is
circular. An observer on Earth, however, has a different experience.
Since the mass ratio does not equal unity, the primary and secondary
stars will occult the light from the other star while at different
displacements from the binary center of mass.  That is, the eclipses as
seen from Earth emanate from different line-of-sight distances. The
secondary eclipses will arrive \textit{later than} the halfway point
between primary eclipses.  (Here we use ``primary eclipse'' to refer to
the eclipse of the primary star, assumed to be brighter and more
massive.)  Using measurements of the orbital period, the sdB velocity
semi-amplitude, the inclination angle, and the time delay (R\o mer delay) of the
secondary eclipse, one can determine both masses, as if the system were
a double-lined spectroscopic binary \citep{kap10}.  The method is
applicable to eccentric systems, too, but the orbital geometry must be
determined first since eccentricity shifts the relative timing of the
two eclipses.

2M 1938+4603, or KIC 9472174 (hereafter, 2M 1938\footnote{The full 2MASS
  designation is 2MASS J19383260+4603591.}), is a bright
(\textit{g}=11.96) eclipsing sdB+dM binary that lies in the
\textit{Kepler} field and a suitable test case for the technique
described above.  The system exhibits grazing primary
and secondary eclipses along with a strong reflection effect from the
cool secondary \citep[hereafter {\O10}]{ost10}; the period is 3.0 h, and
the orbit appears to be nearly circular. The analysis described by
\citet{kap10} predicts the secondary eclipse should occur 2.35 $\pm$
0.10 s after the midpoint between primary eclipses, assuming the mass ratio
from {\O10}'s light curve modeling solution, $q$=0.244 $\pm$ 0.008.  
This time delay should be observable given sufficient
measurements.  After phase-folding a nine day long light curve obtained
from \textit{Kepler} during quarter 0 (Q0) of operations, {\O10} found evidence for more than 55
pulsational frequencies covering both the p- and g-mode domains in frequency space.
Although the amplitudes are quite small ($<$ 0.1 \%), the rich
pulsation spectrum might permit asteroseismic models to establish the
mass of the sdB.  Thus 2M 1938 offers an opportunity to compare mass
determinations for an sdB star, obtained by different methods.

\newpage

Here we present an analysis of more than six months of short-cadence
\textit{Kepler} photometry taken during Q0,
 quarter 5 (Q5)  and quarter 6 (Q6) of operations. Our 
primary goal was to measure the secondary time delay.  Using fits to the
eclipse profiles we confirm that there is a time delay, which we use in
combination with data from {\O10} to compute masses for the sdB and
M dwarf.  We compare our results with the light-curve modeling solution
of {\O10}.

\section{Observations}
\label{sec:obs}
In May 2009 and from March 2010 to September 2010, \textit{Kepler} observed 1436
orbital cycles of 2M 1938 using short-cadence observations.  In this
operating mode, nine 6.02-s exposures, each with a readout time of 0.52
s, are summed into memory to produce an image every 58.85 s (92\% duty
cycle).  We downloaded the public Q0, Q5 and Q6 light curves from the
\textit{Kepler} data archive\footnote{http://archive.stsci.edu/kepler}.
We converted the time stamps, given as Barycentric \textit{Kepler}
Julian Date and accurate to $\pm$ 0.05
s, to Barycentric Julian Date.  \citet{gil10} give additional characteristics of \textit{Kepler}
short-cadence data.

\begin{figure}
\begin{center}
\includegraphics[scale=0.95]{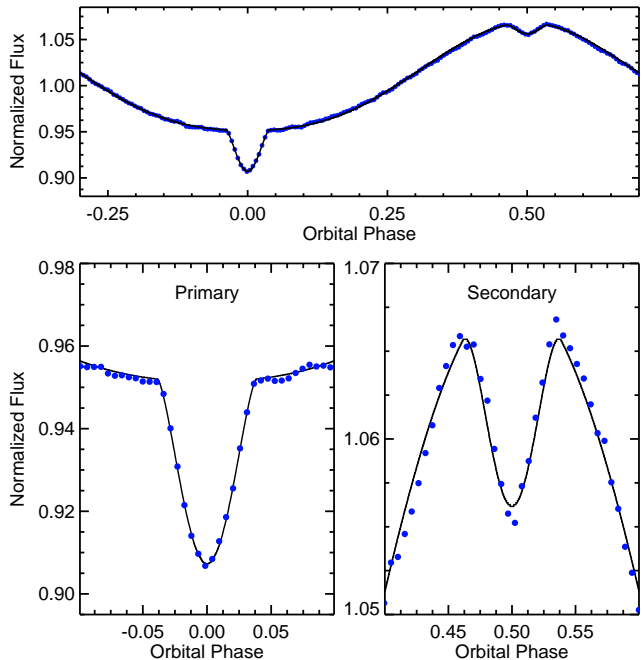}
\caption{\textit{\textbf{Top panel:}} Our best-fitting light curve model 
from {\sc Binary Maker 3.0} (solid black line) plotted on top of a 
randomly-selected cycle in the \textit{Kepler} light curve (blue points).
The secondary eclipse observed during this cycle happens to be particularly
affected by the constructive interference of pulsations and chunking (see text).
The average photometric accuracy in each point is around 0.04\%. 
\textbf{\textit{Bottom panels:}} Enlarged portions of the top panel 
showing the primary eclipse (left) and secondary eclipse (right) 
profiles.  The poor sampling across the eclipses combined with chunking and the pulsations leads to an 
asymmetry in the observed profiles that varies from cycle to cycle.}
\label{fig:model}
\end{center}
\end{figure}

\begin{table}       
\caption{Dates of Primary Eclipse Minimum}
\label{tab:param}
\scriptsize
\begin{center}
\leavevmode
\begin{tabular*}{6cm}{cc} \hline \hline              
BJD & $\sigma$ BJD \\
(d) & (d)\\
\hline
4953.64250  & 0.00003 \\
4953.76824  & 0.00004 \\
4953.89399 & 0.00004 \\
4954.01973 & 0.00004 \\
4954.14553& 0.00004 \\
\hline
\multicolumn{2}{p{.32\textwidth}}{(This table is available in its entirety in a  machine-readable form in the online journal. A portion is shown here for guidance regarding its form and content.)}
\end{tabular*}
\end{center}
  \label{tab:table1}
\end{table}

\begin{table}       
\caption{Dates of Secondary Eclipse Minimum}
\label{tab:param}
\scriptsize
\begin{center}
\leavevmode
\begin{tabular*}{6cm}{cc} \hline \hline              
BJD & $\sigma$ BJD \\
(d) & (d)\\
\hline

4953.57962 & 0.00006\\
4953.70534 & 0.00007\\
4953.83107 & 0.00005\\
4953.95690 & 0.00007\\
4954.08257 & 0.00008\\
\hline
\multicolumn{2}{p{.32\textwidth}}{(This table is available in its entirety in a  machine-readable form in the online journal. A portion is shown here for guidance regarding its form and content.)}
\end{tabular*}
\end{center}
  \label{tab:table2}
\end{table}

\section{Measuring the R\O mer Delay}
\label{sec:eclipse}
We constructed a model light curve to use as a template for fitting
  each eclipse, using {\sc Binary Maker 3.0}\footnote{http://www.binarymaker.com/} with the orbital parameters
  reported by \O 10\footnote{\O 10 do not cite a secondary bolometric albedo; 
we used $A_{\rm 2}$=1.2.}.  As shown in Figure \ref{fig:model},
the model reproduces the observations reasonably well.  
We cross-correlated the primary and secondary eclipse template profiles against
each observed eclipse to determine the times of minima.  The model was re-sampled
step-by-step as it was swept across each eclipse to 
match the sampling of the observations.  Since finite integration times
introduce asymmetric distortions to the eclipse 
profiles  (see \citealt{kip10}), we attempted to emulate 
the exposure times by integrating each sampling point
in a similar manner as the \textit{Kepler} data to better match the observed profiles.  
Before the correlations were made, each eclipse and
its surrounding continuum were cropped from the light curves and
normalized by polynomial fits to the continuum.  The same fit was used
for every model-observation pair.  We note that we do not fit
the eclipse depths or durations in this work; only times of minima are reported.  We also investigated 
the effects of both orbital \citep{sha87} and rotational \citep{gro12} Doppler boosting, which asymmetrically skew the 
eclipse profiles, but found only negligible changes in the fit parameters for 
the expected levels of boosting in this system.

\begin{figure}
\begin{center}
\includegraphics[scale=0.95]{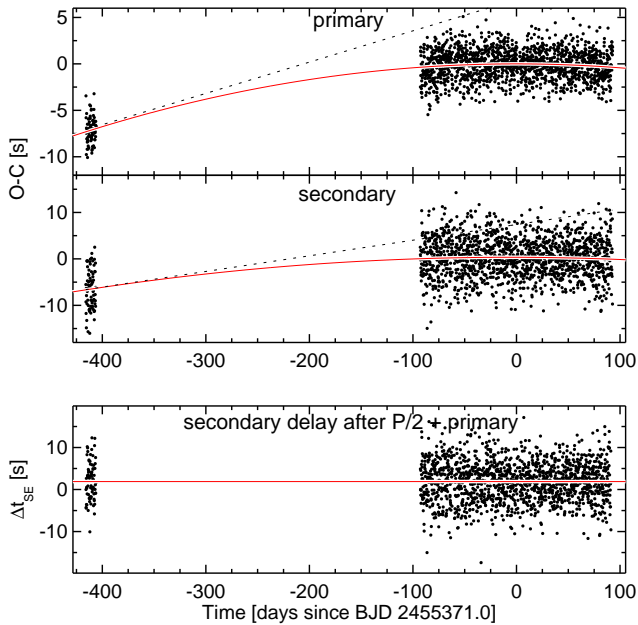}
\caption{\textit{\textbf{Top panels:}} Primary and secondary eclipse O-C
  diagrams (not on the same scale) from the Q0, Q5, and Q6 \textit{Kepler}
  light curves. Both plots show evidence of a period decrease.  The dotted line represents \O10's
  linear ephemeris from Q0, while the solid line shows our quadratic ephemeris.
  \textit{\textbf{Bottom panel:}} Time delay of the secondary eclipse
  with respect to one half period after the primary eclipse, which does
  not significantly change from Q0 to Q5/Q6.  The mean $\Delta t_{\rm SE}$ is shown by a solid horizontal line.}
\label{fig:O-C_full}
\end{center}
\end{figure}

Tables 1 and 2 (full tables available online) present the measured eclipse times, 
which have an average uncertainty of a few seconds.  Using the linear orbital ephemeris
 given by {\O10} as a starting point, we constructed two
{\em observed minus calculated} (O-C) diagrams, one for the primary eclipse
and another for the secondary. These are shown in Figure \ref{fig:O-C_full} and reveal a
 seven-second phase shift between Q0 and the beginning of Q5; the 
data cannot be fitted well with a linear ephemeris.  From a parabolic fit to all three
 quarters, we report an ephemeris for primary eclipses defined by
\begin{equation}
\begin{split}
 & T_{o} = 2455\, 369.422\, 466\, 9 \pm 0.000 \,000\, 5 \,\, {\rm BJD} \nonumber \\
& P =  0.125\, 765\, 251 \pm 0.000\, 000 \,002 \, \,{\rm days} \nonumber \\
& \dot{P} = (-1.23 \pm 0.07) \times 10^{-10} \nonumber \\
\end{split}
\label{eqn:ephemeris}
\end{equation}
The zero-point was chosen to be near the transition time between Q5 and Q6.  
Fits to the secondary eclipse O-C diagram (middle panel of Figure \ref{fig:O-C_full}) give
consistent results for the period and its first derivative, which
corresponds to a nearly 4 ms decrease in the orbital period each year.
The derived period during Q5/Q6 does not agree with the Q0 result of \O 10,
but this is due to the non-zero $\dot{P}$, which was unknown at the time.  Using our updated
ephemeris to derive the period during Q0, we find the same value
they do.  Orbital decays have been observed for other sdB+dM systems (e.g., \citealt{qia12}) and might be 
explained by tidal dissipation, magnetic braking, and gravitational wave emission;
 the latter effect, however, predicts a period change several orders of magnitude slower than what we 
 observe for 2M 1938.   We note that a long-period sinusoidal oscillation in
the O-C diagram could mimic a $\dot{P}$; only additional measurements
over an extended baseline will help discriminate between these possibilities.

 \begin{figure}
\begin{center}
\includegraphics{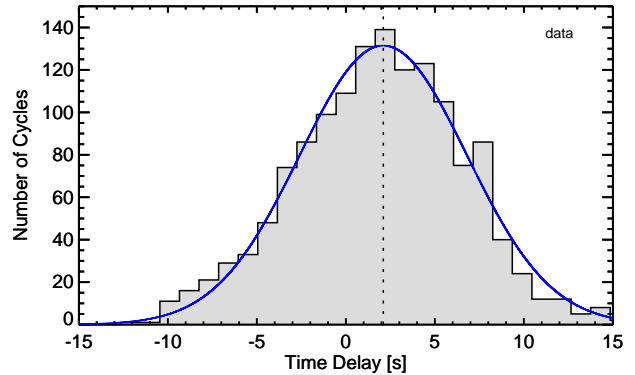}
\caption{Histogram of the secondary eclipse time delays, 
measured with respect to the mid-point between primary 
eclipses.  Data represent 1493 pairs of primary and secondary 
eclipses obtained in Q0, Q5, and Q6. The Gaussian (solid line) 
that best fits the distribution is centered at 2.06 $\pm$ 0.12 s.}
\label{fig:histogram_obs}
\end{center}
\end{figure}

\begin{figure*}
\begin{center}
\includegraphics{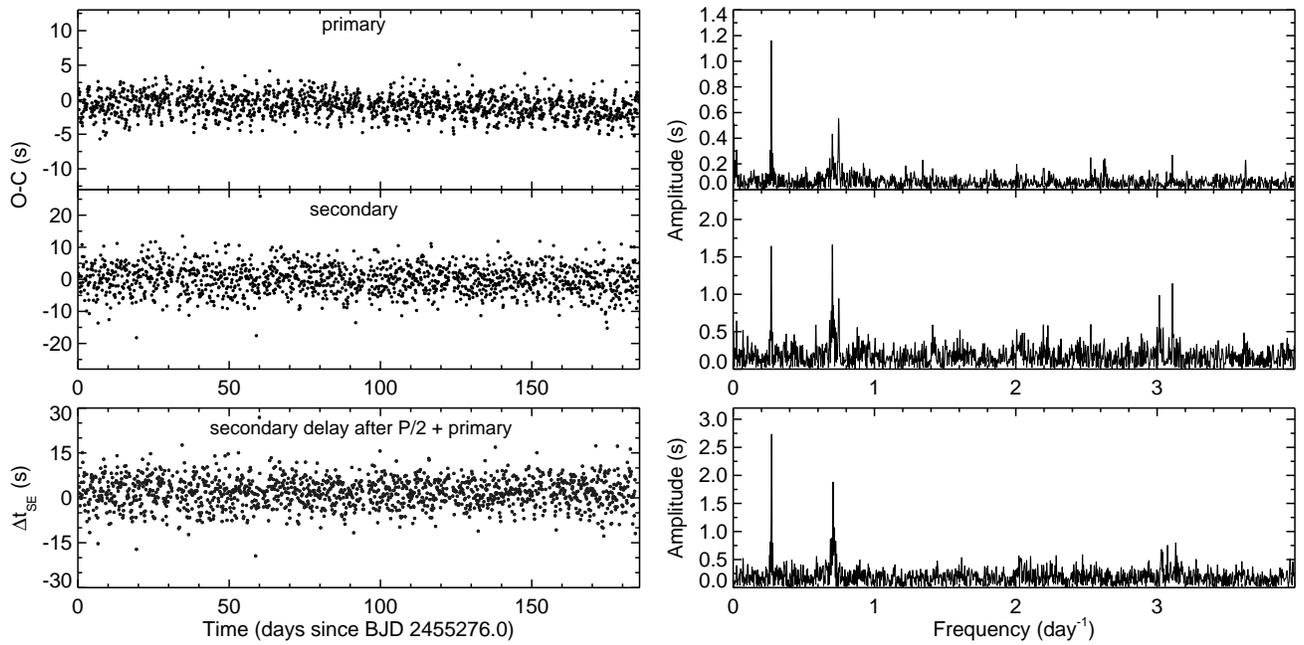}
\caption{ \textit{\textbf{Top panels:}} O-C diagrams (left) constructed
  from primary and secondary eclipse timings in Q5/Q6 and their
  respective Fourier transforms (right).  Several periodic signals are visible 
  in the eclipse timings with periods ranging from 0.25-3.7 days  \textit{\textbf{Bottom
      panels:}} Time delay of the secondary eclipse with respect to one
  half period after the primary eclipse (left) and its FT (right).  Some of the same 
  periodicities detected in the O-C diagrams are also present here.}
\label{fig:O-C_FT_obs}
\end{center}
\end{figure*}

\begin{figure*}
\begin{center}
\includegraphics{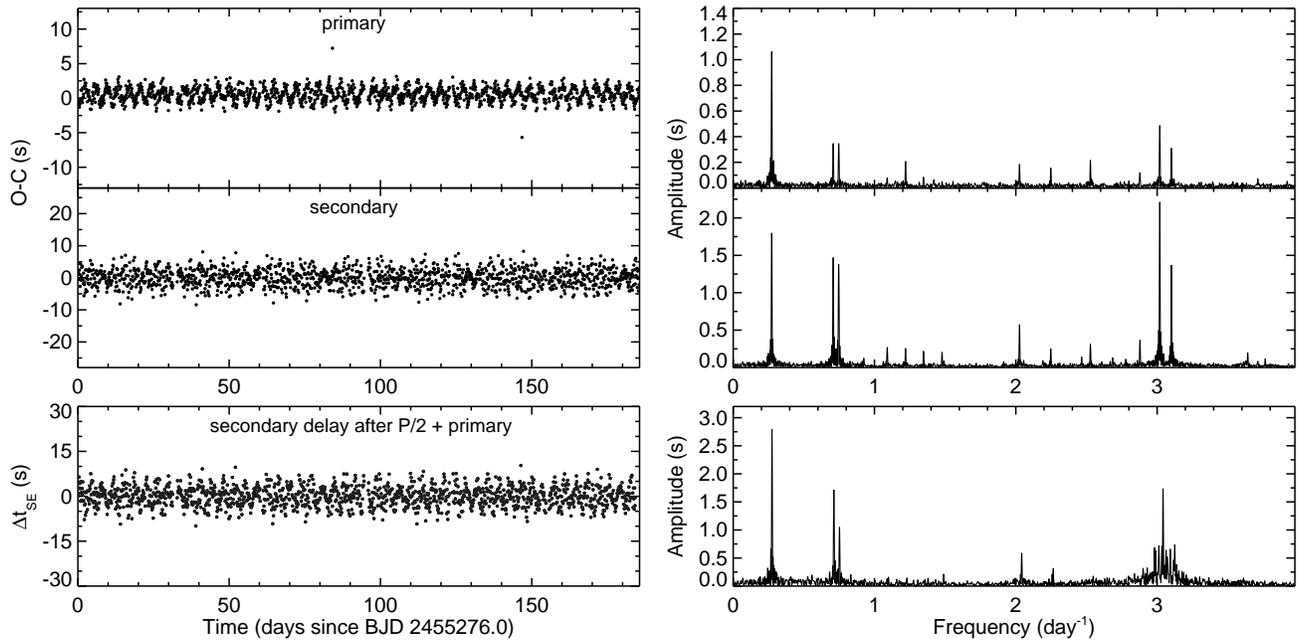}
\caption{Same as Figure \ref{fig:O-C_FT_obs}, but for a synthesized 
binary light curve with all 55 pulsation frequencies from \O 10 added.  
The pulsations and finite sampling skew the eclipse profiles in a 
periodic fashion that gives rise to apparent oscillations in the 
arrival times of the eclipses.  Note the alignment between these 
signals and those in Figure \ref{fig:O-C_FT_obs}.  No noise was 
added to the model light curve. }
\label{fig:O-C_FT_model}
\end{center}
\end{figure*}

As mentioned above, we do not expect the secondary eclipse to occur exactly halfway
between primary eclipses since the mass ratio given by {\O10} is far from unity.  
In the bottom panel of Figure \ref{fig:O-C_full}, we plot
deviations of the secondary eclipse times from 1/2 period after the
primary eclipses (hereafter, $\Delta t_{\rm SE}$).  Even with the large spread
in $\Delta t_{\rm SE}$ values, the mean is visibly offset from zero.  The offset becomes
clearer in a histogram of the measured time delays, as shown 
in Figure \ref{fig:histogram_obs}.  A Gaussian centered at 2.06 $\pm$ 0.12 s 
with a full width half maximum (FWHM) of 11 s  fits the distribution well.  Its centroid
agrees with the mean of all $\Delta t_{\rm SE}$ values, 1.88 $\pm$ 0.13 s.  
\textit{Indeed, the secondary eclipse lags behind the mid point between 
primary eclipses}, by approximately two seconds.

Before proceeding, we computed the Fourier transforms (FTs) 
of the O-C and $\Delta t_{\rm SE}$ curves to identify and analyze any
periodic signals that might be present.  Only Q5 and Q6 were considered in this exercise 
since the addition of Q0 and the gap it introduces complicates the window function.  
The upper panels
of Figure \ref{fig:O-C_FT_obs} present the O-C diagrams next to
their amplitude spectra, which are plotted out to the Nyquist frequency
(3.976 day$^{-1}$ = 1/2P).  The bottom panels shows the same diagrams
for the secondary eclipse delay.
 
Several strong periodic signals are present in the eclipse timings, all of which arise 
from the rapid pulsations of the sdB star and inadequate emulation of the integration time
of the observations.  The latter issue, which we refer to as a `chunking' problem, stems from 
the finite sampling time of each \textit{Kepler} point, 
which distorts the light curve (see \citealt{kip10}).  Even though 
we attempted to avoid the effects of chunking by 
taking integration into account in our model template,
small mismatches in the way our model and data were binned remain.  The same problem
in accounting for chunking was encountered by \citet{kip11}. The effects of
the sdB pulsations on the eclipse timings are even more pronounced.  
Although none of the 55 pulsational frequencies
detected by \O 10 has an amplitude greater than 0.05\%, the \textit{Kepler} data are so precise
these small oscillations also introduce noticeable asymmetries into the eclipse profiles.   
The phasing of the distortions changes from cycle to cycle, pulling around 
the best-fitting centroid values on timescales related to the beating of the pulsations
with the orbital period.  In some cases, such as the 
$f$=0.26 d$^{-1}$ signal, the peak-to-peak amplitude
about the mean approaches several seconds.

To model the influence of pulsations and chunking on the eclipse timings, we 
constructed a fake, noiseless light curve using our {\sc binary maker 3.0} template 
with the same sampling and integration as the Q5/Q6 $\textit{Kepler}$ data.  We included all of the 
known pulsations to the light curve, using the frequencies and 
amplitudes given in Table D1 of \O 10.  Their phases were
defined randomly since they were not provided by \O 10.
We forced the secondary eclipse
 to occur exactly halfway between primary eclipses to determine whether the pulsations affect
 our measurement of the mean time delay.
A second model light curve, identical to the first sans pulsations, was also constructed.  
We repeated the entire analysis procedure for the synthesized data; 
Figure \ref{fig:O-C_FT_model} shows the resulting eclipse measurements
and their amplitude spectra.

\begin{figure}
\begin{center}
\includegraphics{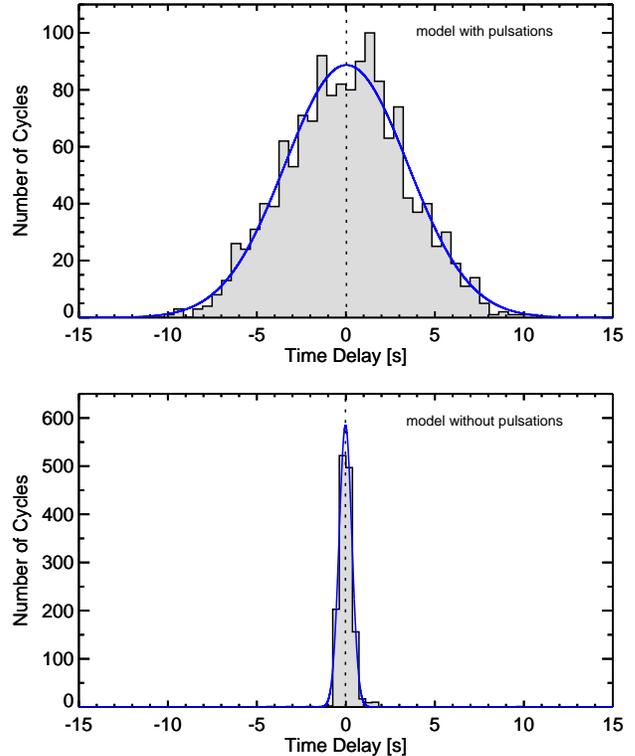}
\caption{Same as Figure \ref{fig:histogram_obs}, but for a noiseless, synthesized 
light curve with (top) and without (bottom) pulsations.  The model was 
constructed with secondary eclipses occurring exactly halfway between 
primary eclipses.  The presence of the sdB's pulsations inflates the 
distribution of eclipse measurements by a factor of 9.7 but does not 
change the mean significantly.}
\label{fig:histogram_model}
\end{center}
\end{figure}

The influence of pulsations and finite sampling on the eclipse timings is drastic; the O-C diagram FTs
reveal a large number of periodicities, none of which represents a `real' signal in the data.
A comparison of Figures \ref{fig:O-C_FT_obs} and \ref{fig:O-C_FT_model} 
shows that each frequency in the observed timings corresponds to one of the spurious
signals generated by adding the finite sampling and the sdB pulsations to our model light curve.  
Most of the signals arise from the stellar pulsations. Their presence, however,
does not offset the mean delay measured for the secondary eclipse.  The best-fitting Gaussian
to the distribution of $\Delta t_{\rm SE}$ values, 
shown in the top panel of Figure \ref{fig:histogram_model}, is centered 
at 0.03 $\pm$ 0.11 s, consistent with the modeled offset of zero.  
Luckily, the spurious timing 
oscillations are so rapid compared to the run length that we sampled
an adequate number of cycles during Q5/Q6 that their mean is very nearly zero.  
For this reason, the uncertainties of the pulsation phases do not affect our 
results.  If we repeat the above analysis using the model without pulsations, the distribution 
is significantly narrower, as seen in the bottom panel of Figure \ref{fig:histogram_model}.  
The best-fitting Gaussian centroid in this case is -0.023 $\pm$ 0.003 s.
Although the pulsations don't significantly affect the mean time delay measured, they do
 inflate the distribution width by a factor of 9.7.

Our light-curve synthesis exercise tells us the centroid of the best-fitting Gaussian to
the observed $\Delta t_{\rm SE}$ distribution in Figure \ref{fig:histogram_obs} 
reflects the true offset in the system, even though the 
distribution of measurements is inflated significantly by the pulsations.
Thus, we report a time delay of $\Delta t_{\rm SE} = 2.06 \pm 0.12 s$ for the secondary
eclipse, as measured from the mid point between primary eclipses. 
Note that this offset remains constant from Q0 to Q5/Q6 (bottom panel of
Figure \ref{fig:O-C_full}), in spite of the seven-second phase shift from the period change.

\section{Derivation of the Masses}

We can derive the masses of the hot subdwarf ($M_{\rm sd}$) and cool
companion ($M_{\rm c}$) from the eclipse timings using the equations
presented by \citet{kap10}.  In a binary system with exactly circular orbits and
unequal masses, secondary eclipses will not occur exactly 1/2 period
after the primary eclipses due to the extra light-travel time (LTT)
required. This R\o mer delay ($\Delta t_{\rm LTT}$) is given by

\begin{equation}
\Delta t_{\rm LTT} = \frac{PK_{\rm sd}}{\pi c}\left( \frac{1}{q}-1\right)
\label{eqn:deltat}
\end{equation}\\
where $P$ is the orbital period, $K_{\rm sd}$ the sdB orbital velocity, $q$
the mass ratio (M$_{\rm c}$/M$_{\rm sd}$).  Small eccentricities will affect the
relative timing of the primary and secondary eclipses through an
additive\footnote{Eqn. \ref{eqn:deltat} should also be multiplied by an
  eccentricity term, but this addition changes $\Delta t_{\rm LTT}$ by less
  than 1\% for $e < 0.1$.  We ignore it here.} term
$\Delta t_e $ defined by
\begin{equation}
\Delta t_e \simeq \frac{2Pe}{\pi} \cos  \omega
\label{eqn:delta_e}
\end{equation}
where $e$ is the eccentricity and $\omega$ the argument of periapsis
(\citealt{ste40,win10,kap10}\footnote{The \citet{kap10} expression (his Eqn. 6) 
is missing a factor of 2.}). Thus, for small eccentricities, the total shift of the
secondary eclipse with respect to 1/2 orbital period after the primary
eclipse is
\begin{equation}
\Delta t_{\rm SE} \simeq \Delta t_{\rm LTT} + \Delta t_e
\label{eqn:SE}
\end{equation}
In \S \ref{sec:eclipse}, we found an average secondary eclipse time delay of
about two seconds with respect to the mid-point between primary
eclipses.  This measurement represents $\Delta t_{\rm SE}$ in the above
equation, and in order to use it to calculate the masses, the
contribution from the eccentricity  ($\Delta t_{\rm e}$) must be identified.   
Surprisingly, an eccentricity of only $e=0.0003$ could 
 shift the secondary eclipse by the full amount we observe; to separate the R\o mer delay
  out from the total measured delayed requires knowing
  the system's eccentricity to a level of precision better than this.  The radial velocity curve
  published by \O 10 only limits $e$ to $\sim0.02$ or smaller\footnote{as estimated 
  from an F-test showing that e $>$ 0.02 gives a significantly worse fit than a circular-orbit solution.}.  Theoretically, one 
  could measure the eccentricity by looking for the apsidal
  motion predicted by general relativity and classical mechanics, which is straight-forward
  to calculate.  The periastron advance (changing $\omega$) induced by these effects
  would give rise to oscillations in the eclipse timings, and in
  particular $\Delta t_{\rm SE}$, with known periods. The presence or absence
  of such a signal in the data allows one to place upper limits on the eccentricity.
 For typical eclipsing sdB+dM systems, unfortunately, the expected precessional 
 periods are several decades, too long to be measured
  with the current dataset. 

Without knowing the eccentricity to the required precision, we continue 
under the assumption of a circular orbit.  Combining
the observed 2.06-s delay with $P$ and $K_{\rm sd}$ (from {\O10}), we derive
a mass ratio (via Eqn. \ref{eqn:deltat}) of $q=0.2691 \pm0.0018$.  This
result is independent of the orbital inclination angle.
Upon assuming a particular inclination, we can also solve for the
individual masses by combining Eqn.\ \ref{eqn:deltat} with Kepler's
Third Law, as done by \citet{kap10} in his Eqn.\ 7.  He 
assumes a perfectly edge-on system in deriving
this expression, and so it must be modified by a multiplicative term (sin
$i$)$^{-3}$ for our use.  If we adopt the light curve modeling result $i$=69.45
$\pm$ 0.02 $\deg$ from {\O10}, we derive masses of M$_1$=0.372 $\pm$
0.024 M$_{\sun}$ and M$_2$=0.1002 $\pm$ 0.0065 M$_{\sun}$.  Additional
orbital parameters calculated from the timing method are summarized in
Table \ref{tab:system_parameters}.

Our derived mass ratio does not agree with the light-curve modeling
  results of {\O10}, from which we infer q=0.244 $\pm$ 0.008; the disagreement in the masses themselves is even more
  pronounced.  Their mass ratio predicts
 a time delay of 2.35 $\pm$ 0.10 s, which is 0.29 $\pm$ 0.16 s longer than our result.  This difference (significant at roughly 2$\sigma$), can easily be explained
 if the eccentricity is as small as $e=0.00004 \pm 0.00002$!  We cannot measure 
 such a minuscule departure from the circular-orbit 
 case using the currently-available data.  Although
 non-zero $e$ likely explains the disagreement, shortcomings in the light curve 
 modeling might also be at play.  Presumably, a revised light-curve
analysis with $q$ fixed at $0.2691\pm0.0018$ would result in a slightly
different inclination.  Unfortunately, the brief description of the light-curve modeling given in
{\O10} does not allow us to predict a revised inclination with confidence.
Our inferred masses are therefore provisional. We note, however, that
in order to get an sdB mass equal to
0.48 M$_{\sun}$ ({\O10}'s derived value) using the observed $P$, $\Delta
t_{\rm LTT}$, and $K_{\rm sd}$, the inclination would have to be 66.7 $\deg$, 
in disagreement with the light curve modeling result. This discrepancy might be explained by a
number of uncertainties associated with the light curve solution,
including how to accurately model the M dwarf albedo and the \textit{Kepler}
bandpass.  Their solution also depends greatly on the quoted value of log
\textit{g}.  Significant changes larger than their 1-$\sigma$ error
(0.009 dex) might come about from non-LTE modeling (versus their LTE
solution), unaccounted--for rotational broadening of the H Balmer lines,
and incomplete removal of the orbital velocities upon summation of the
spectra for fitting.  In light of some of these effects, \O 10
themselves stress that their model parameters are subject to systematic
errors that might be larger than the statistical errors they cite.

\setcounter{table}{2}

\begin{table*}       
\caption{System Parameters} 
\label{tab:param}
\scriptsize
\begin{center}
\leavevmode
\begin{tabular}{lllll} \hline \hline              
Param & Value & Error & Unit & Comments \\
\hline
$P$ & 3.018\,366\,023& $\pm$ 0.000\,000\,048& hrs & measured from \textit{Kepler} Q0+Q5+Q6 primary minima at start of Q6\\
$\dot{P}$ & $-$1.23 & $\pm$ 0.07 & 10$^{-10}$ s s$^{-1}$ & measured from \textit{Kepler} Q0+Q5+Q6 primary minima\\
$K_{\rm sd}$ & 65.7 & $\pm$ 0.6 & km s$^{-1}$ & measured by \citet{ost10}\\
$i$ & 69.45 & $\pm$ 0.02 & deg & calculated by \cite{ost10}\\
$\Delta t_{\rm SE}$ & 2.06 & $\pm$ 0.12 & s & measured from \textit{Kepler} Q0+Q5+Q6 (from Gaussian fit to distribution) \\
\hline
$q$ & 0.2691$^a$ & $\pm$ 0.0018 &-- & calculated from $\Delta t_{\rm SE}$, P, K$_{\rm sd}$\\
$M_{\rm sd}$ & 0.372$^a$ & $\pm$ 0.024 & M$_{\sun}$ & calculated from $\Delta t_{\rm SE}$, P, K$_{\rm sd}$, sin $i$\\
$M_{\rm c}$ & 0.1001$^a$ & $\pm$ 0.0065& M$_{\sun}$ & calculated from $\Delta t_{\rm SE}$, P, K$_{\rm sd}$, sin $i$\\
$a$ & 0.823$^a$ & $\pm$  0.015 &  R$_{\sun}$ & calculated from P, K$_{\rm sd}$, M$_{\rm sd}$+M$_{\rm c}$\\
\hline
$R_{\rm sd}$ & 0.196$^a$ & $\pm$ 0.049 & R$_{\sun}$ & estimated using log \textit{g} and M$_{\rm sd}$\\
$K_{\rm c}$ & 244.2$^a$  & $\pm$ 2.8 & km s$^{-1}$ & estimated using $q$ and $K_{\rm sd}$ \\
\hline
\multicolumn{5}{l}{$^a$assumes a circular orbit.}
\\
\\
\end{tabular}
\end{center}
  \label{tab:system_parameters}
\end{table*}

\vspace{10mm}
\section{Conclusions}

Using three months of short-cadence \textit{Kepler} photometry, we have
shown that primary and secondary eclipse timings can help constrain the
component masses in sdB+dM eclipsing binaries using the technique
described by \citet{kap10}. The secondary eclipses of 2M 1938 arrive
nearly two seconds after the halfway point between primary eclipses.  
Assuming a circular orbit,  we are able to derive the
mass ratio and individual component masses from this delay, the orbital period,
the sdB velocity, and the inclination angle.  Our total system mass and the
mass ratio disagree with the results of {\O10}. However, this difference may be
reconciled if the system has an eccentricity no less than $\sim$0.00004, which is too small
to be measured using the currently-available data.  This work, and the concurrent
study of KOI-74 by \citet{blo12}, represent the first detections of 
secondary eclipse R\o mer delays
in compact binaries.  The results of both studies, however, are fundamentally \textit{inconclusive}
 at this time; until the eccentricities in these systems are determined to 
 a precision of $\sim$ 10$^{-5}$, it is impossible to separate out the R\o mer effect
contribution from the total observed time delay.  An
 eccentricity-induced time delay can easily and naturally explain
the apparent disagreement between the R\o mer delay
and binary light curve modeling solutions for both 2M 1938+4603 and KOI-74.

If the binary light curve modeling solution and 
derived mass ratio from \O 10 are accurate, the time delay we measure
implies 2M 1938 has a non-circular orbit with eccentricity no smaller than
$e\sim0.00004$.  Close sdB+dM binaries are generally expected to have
circular orbits, as they are likely the products of CE 
evolution.  Even if some level of eccentricity remains after their
formation, their circularization timescales were thought to be short
enough for the orbits to circularize within the lifetime of the sdB, around 100 Myr \citep{tas88,zah77}.    
Several phenomena might explain a small eccentricity in 2M 1938' s orbit.
Some recent studies of sdB+dM systems (e.g., \citealt{pab12}) show the synchronization 
timescales might be longer than the EHB lifetime.   The system also might
 have been ÒbornÓ (following the CE phase) with the M dwarf's spin 
 misaligned with the orbit, since in the wider pre-CE binary there was 
 no reason for it to become aligned. Spin-orbit misalignment has been 
 reported for the young eclipsing binary DI Her \citep{alb09}. 
 The resultant torques will give rise to a time-varying eccentricity, apsidal 
 motion, precession of the orbital plane, and spin precession. (In this 
 case the misaligned figure of the M dwarf would probably vitiate some 
 of the assumptions of the light-curve model used by \O10.) A third body 
 orbiting outside the 3.0 h binary would perturb any initially circular orbit, 
 resulting in orbital eccentricity. If the third-body orbit were not coplanar 
 with the inner binary, one would expect to observe precession of the inner 
 binary's orbital plane as well. Any eccentricity will be damped out, at the 
 expense of orbital energy, so another prediction is that the orbital period 
 should be decreasing with time (as observed, although a longer timebase 
 is needed to be sure that this is a truly secular phenomenon, rather than 
 periodic or quasi-periodic). \citet{egg06} discusses these processes 
 and their associated amplitudes and timescales extensively. Random 
 convective motions in the M dwarf could also lead to an eccentric orbit, 
 since they induce fluctuations in the exterior gravitational field \citep{phi92,lan01}.

Due to strong irradiation,
we expect some level of H-alpha emission from the M dwarf.  If such a
feature is detected and the eccentricity of the system is eventually determined
to adequate precision, 2M 1938 will be unique among binaries in that the
individual masses can be computed using at least four different techniques:
double-lined spectroscopic binary analyses, asteroseismology, 
eclipsing binary light curve modeling, and eclipse timing monitoring.
Comparisons of these results will help constrain the underlying
physics in the light-curve modeling and asteroseismic codes.

\begin{acknowledgements}
  This material is based upon work supported by the National Science
  Foundation under Grant No.\ AST-0908642.  We are grateful to 
  an anonymous referee for particularly helpful comments and suggestions
  that greatly improved this manuscript.
The data presented in this
  paper were obtained from the Multimission Archive at the Space
  Telescope Science Institute (MAST). STScI is operated by the
  Association of Universities for Research in Astronomy, Inc., under
  NASA contract NAS5-26555. Support for MAST for non-HST data is
  provided by the NASA Office of Space Science via grant NNX09AF08G and
  by other grants and contracts.  This paper includes data collected by
  the Kepler mission. Funding for the Kepler mission is provided by the
  NASA Science Mission directorate.
\end{acknowledgements}

%
%

\end{document}